\documentstyle[preprint,aps]{revtex}
\begin{document}
\newcommand\beq{\begin{equation}}
\newcommand\eeq{\end{equation}}
\newcommand\bea{\begin{eqnarray}}
\newcommand\eea{\end{eqnarray}}

\draft

\title{A Fermion-like description of condensed Bosons in a trap}
\author{R.K.Bhaduri and M. V. N. Murthy\thanks{Permanent Address: The 
Institute of Mathematical Sciences, Madras 600 113, India}}
\address
{Department of Physics and Astronomy, McMaster University, \\
Hamilton, Ontario, Canada L8S 4M1} 
\date{\today}
\maketitle
\begin{abstract}
A Bose-Einstein condensate of atoms, trapped in an axially symmetric harmonic
potential, is considered. By averaging the spatial density along the
symmetry direction over a length that preserves 
the aspect ratio, the system may be mapped on to a zero 
temperature noninteracting Fermi-like gas. The ``mock fermions'' have a 
state occupancy factor $(>>1)$ proportional to the ratio of the coherance 
length to the ``hard-core'' radius of the atom. The mapping reproduces the 
ground 
state properties of the condensate, and is used to estimate the vortex 
excitation energy analytically. The ``mock-fermion'' description predicts 
some novel collective excitation in the condensed phase. 
\end{abstract} 
\vskip 1 true cm

\narrowtext

Recently there has been a renewed interest in the Bose-Einstein
condensation(BEC) of a gas after its experimental
demonstration\cite{anderson} with rubidium vapour in a trap at a
temperature of 170 nanokelvin and at a number density $\rho = 2.5\times
10^{12}$ atoms per cc.  This experiment has been followed by others using
alkali atoms\cite{others}. At these temperatures the atoms form a weakly
interacting metastable gas of Bosons.  For a non-technical account see the
review by Burnett\cite{burnett}.  
In a typical device, atoms are trapped in a potential
which is well described by an axially symmetric parabolic confinement. The
oscillator frequency in the symmetry direction is larger than the
frequency in the plane perpendicular to it. 
The experimental situation of interest
to us is the one with rubidium vapour, where the s-wave scattering length
between two atoms is known to be positive. 
The effect of the  
interatomic interaction may be mocked up by a repulsive 
pseudo-potential\cite{yang}. The interaction energy is propotional to 
$a\rho^2$, where $a$ is the s-wave scattering length and $\rho$ is the
number density of the atoms. 
The properties of the condensate have been studied by constructing 
the density functional
involving this replusive interaction energy, and the potential energy of the 
atoms in the trap \cite{baym,stringari,fetter}. 
In this paper, we first note that by averaging the spatial density
of the condensed bosons along one direction, it may be reduced to the 
same form as the density of a non-interacting Fermi gas.
We chose the averaging direction to be the symmetry axis (the 
z-direction ), along which 
the harmonic confinement is steeper. {\it This enables us to use the Fermi gas 
model to compute the low-lying planar excitations of the 
condensate in the shallow well}. Moreover, the averaging distance is chosen 
to preserve the aspect ratio (the ratio of the length scales in the planar to 
perpendicular directions ) of the original trap. The Bose-condensate is now 
described by a three-dimensional noninteracting ``Fermi'' gas, trapped in the 
same planar parabolic potential as the original system, but free to move 
along the z-direction within the averaging distance. 
One peculiarity of these mock-fermions, as we call them, is their occupancy 
factor per state. Instead of being one (or zero ) at $T=0$, it is multiplied 
by a large factor proportional to $\lambda_F/2a$, where $\lambda_F=h/p_F$, 
and $2a$ the apparent size of the atom. In fact, the Fermi momentum 
$\hbar k_F$ is such that $k_F^{-1}=\xi$, where $\xi=(8 \pi \rho a)^{-1/2}$ is 
just the coherance length in the bose condensate.
It is remarkable that the kinetic energy of these mock fermions exactly 
reproduces the condensate energy in the large-N limit. The latter is in fact 
calculated by neglecting the kinetic energy of the bosons.
After having shown this equivalence, we go on to use this model to calculate 
some other 
properties. These include the velocity of sound, and the vortex excitation 
energy. The sound velocity in the mock-fermion ideal gas is the same as in 
the Bose-condensate. A simple estimate of the vortex excitation energy is 
made by ``digging'' a hole in the central density, i.e., 
promoting all the s-state 
mock-fermions out of the Fermi sea to states of non-zero angular momentum. 
This reproduces the numerical results of Dalfovo and 
Stringari\cite{stringari} satisfactorily.
The latter calculation involved solving a nonlinear Schr\"{o}dinger 
equation that was obtained from the density-functional formalism.
Finally, our description also predicts some novel collective excited states 
with zero angular momentum involving a large number of mock-fermions.

~~~~We begin with the ground state energy for condensed bosons given by the 
Ginzberg-Gross-Pitaevskii\cite{functional} energy functional,
\beq
E[\psi] = \int d^3r \left[ \frac{\hbar^2}{2m} |\nabla \psi|^2 +\frac{m}{2} 
(\omega_{\perp}^2 r_{\perp}^2 + \omega_3^2 z^2)|\psi(r)|^2 + \frac{2\pi 
\hbar^2 a}{m} |\psi(r)|^4\right], \label{epsi}
\eeq
where $m$ is the mass of the atom, $\omega_{\perp}, \omega_{3}$ denote 
the oscillator frequencies in the transverse direction and in the 
direction of the symmetry axis( z-axis), and $a$ is the s-wave scattering 
length which defines the strength of the interaction in the 
pseudo-potential method. The condensate wave function is usually denoted 
as 
\beq
\psi(r) = \sqrt{\rho(r)} \exp{is(r)}.
\eeq
We work in 
the limit of strong repulsive interaction where the kinetic energy term 
can be neglected in the condensate phase. In the large N-limit we can 
then obtain the density $\rho(r)$ by minimising $(E-\mu N)$, where 
$\mu$ is the chemical potential. This is the 
Thomas-Fermi expression for the density within the classical turning 
points\cite{stringari} : \beq
\rho(r) = \frac{m}{4\pi \hbar^2 a }[\mu -\frac{m}{2}(\omega_{\perp}^2 
r_{\perp}^2 +\omega_3^2 z^2)]. \label{rho}
\eeq
The density is zero outside the turning points.  
Note that when the scattering length $a$ is positive, the chemical 
potential $\mu$ is necessarily positive since $\rho(r) \geq 0$. This is 
the situation with the BEC of rubidium atoms. It is now straight-forward to 
calculate the particle number and the 
energy using the density given above. The particle number is 
given by,
\beq
N = \int d^3 r \rho(r) = \frac{a_3^2}{15 a a_{\perp}}(\frac{2\mu}{\hbar 
\omega_{\perp}})^{5/2}, \label{number}
\eeq
where 
\beq
a_{\perp} = \sqrt{\frac{\hbar}{m\omega_{\perp}}};~~~~~~
a_{3} = \sqrt{\frac{\hbar}{m\omega_{3}}},
\eeq
The integration limits are set by the turning points in the $z$- and the 
$r_{\perp}$-directions. These may be obtained by first performing the $z$-
integration for a fixed $r_{\perp}$, and then allowing the latter to vary 
within the prescribed limits.
The condensate energy is obtained  by substituting the expression~(\ref{rho}) 
for the density in Eq.(\ref{epsi}), and is given by
\beq
E = \frac{a_3^2}{42 a 
a_{\perp}}~(\frac{2\mu}{\hbar\omega_{\perp}})^{7/2}~\hbar\omega_{\perp}. 
\label{energy} \eeq
As noted before, the kinetic energy term in Eq.~(\ref{epsi}) is neglected.
These are known results. 
Before we go further, we note that the ratio $E/N = 5\mu/7$. This is 
indeed the energy per particle in a non-interacting fermionic system 
whose single-particle density of 
states $g(E)\propto  E^{3/2}$, obtained by convolving the state density 
in a planar parabolic 
potential, with the density in the transverse direction.

~~~~We now take the crucial step by averaging the density 
in the direction of the symmetry axis over the a scale 
$L_3$ : 
\beq
\rho_{a}(r_{\perp}) = {\frac{1}{L_3}}\int dz~ \rho(r)\;.
\eeq
We refer to $\rho_a$ as the mock-fermion density for reasons that will be 
clear soon. The length scale $L_3$ ($-L_3/2 
\leq z \leq L_3/2$) will be fixed later by fitting the aspect ratio. A little 
algebra shows that the result may be written as
\beq
\rho_{a}(r_{\perp})=\frac{1}{\alpha}\rho_{TF}~(r_{\perp}), \label{rhofermi}
\eeq
where
\beq
\rho_{TF}(r_{\perp}) = \frac{1}{6\pi^2}~\left[~\frac{2m}{\hbar^2} 
~(\mu-\frac{m\omega_{\perp}^2}{2} r_{\perp}^2)~\right]^{3/2},
\label{tf}
\eeq
and the dimensionless parameter $\alpha$ is given by 
\beq
\alpha = \frac{L_3 a}{\pi a_3^2}.
\label{hal}
\eeq
The Thomas-Fermi expression for the density, $\rho_{TF}$, given by 
Eq.~(\ref{tf}), is in fact the density of spinless fermions confined in a 
harmonic potential in the plane and free to move in the z-direction. 
Note from Eqs.~(\ref{rhofermi}, \ref{hal}) that for 
$a_3 \rightarrow 0$, the system becomes two-dimensional, and there is no 
condensation since $\rho_a \rightarrow 0$. Similarly, the condensation 
density is depleted as the scattering length $a$ is increased. Note that
$1/\alpha$ plays the role of the occupancy factor which is not necessarily 
unity. It is in this sense we call these particles mock-fermions.

~~~To proceed further we first compute the number of particles $N$ 
and the condensate energy $E$, using $\rho_a (r_{\perp})$ as given by 
Eq.~(\ref{rhofermi}). Once again the number of particles is defined by
\beq
N = L_3\int d^2 r_{\perp} \rho_a(r_{\perp}) 
=\frac{L_3}{15 \pi \alpha a_{\perp}}(\frac{2\mu}{\hbar 
\omega_{\perp}})^{5/2}. \label{newnumber}
\eeq
Substituting for $\alpha$ from Eq.(\ref{hal}) reproduces the particle 
number 
given by Eq.(\ref{number}). 
The total energy in fermion-like picture is given by,
\beq
E[\rho_a] = \frac{1}{\alpha} \int d^3r \left[ 
\frac{\hbar^2}{2m} \tau (\vec r) + U(\vec 
r)\rho_{TF}(\vec r)\right],\label{newen} \eeq
where $\tau(\vec r)$ is the kinetic energy density of the fermions in the 
Thomas-Fermi approximation, 
\beq
\tau(\vec r) = \frac{1}{10\pi^2} (6\pi^2 \rho_{TF})^{5/3}
\eeq
and $U(\vec r)$ is the confining oscillator potential in the plane. The 
particles are 
of course free to move in the z-direction, but confined within the 
length $L_3$. 
Substituting for the density and the kinetic energy density in 
Eq.(\ref{newen}), we immediately obtain the total energy 
\beq
E = \frac{L_3}{42\pi\alpha a_{\perp}}
~(\frac{2\mu}{\hbar\omega_{\perp}})^{7/2}~\hbar\omega_{\perp}.  
\label{newenergy} \
\eeq
Again, substituting for $\alpha$ yields the energy of the Bose condensate 
as given by Eq.(\ref{energy}).  This is remarkable since the energy of 
the interacting bosons in the BEC is identical to that of the 
non-interacting mock-fermions.

Note, from Eqs.~(\ref{newnumber}, \ref{newenergy}), that both the particle 
number $N$ and energy $E$ are independent of the 
choice of $\L_3$. This is because it 
comes in the combination $L_3/\alpha$. This does not mean that the choice 
of $L_3$ may be arbitrary, since we shall presently see that that the excited 
state properties depend crucially on it.  
A physically meaningful 
way of fixing the length is to set it through the aspect ratio 
$\sqrt{<x^2>/<z^2>}=\omega_3/\omega_{\perp}$. Consequently
the size of the cloud is approximately the same in BEC and in the 
mock-fermion picture. We have cylindrical symmetry in 
the latter instead of ellipsoidal symmetry 
of the BEC. In the fermion picture $<z^2> = L_3^2/12$. The average 
distance in the plane is  
($<r_{\perp}^2> =\frac{2 a_{\perp}^2 \mu}{7\hbar\omega_{\perp}}$) is 
the same as in the BEC.  Equating the aspect ratios in both pictures we get,
\beq
L_3 = \sqrt{\frac{24\mu}{7m\omega_3^2}}~~~.
\eeq 
This then determines the occupancy factor $\alpha$ uniquely,
\beq
\alpha = \frac{a}{\pi} \sqrt{\frac{12}{7}} \sqrt{\frac{2m\mu}{\hbar^2}}~~~,
\label{pond}
\eeq
where $a$ is the scattering length.
Note that $\alpha$ now depends on the chemical potential. 
To obtain the sound velocity in the interior of the cloud, we set the 
density $\rho_a$ from Eq. (\ref{rhofermi}) to its central value:
\begin{equation} 
\rho_a = \frac{1}{6\pi^2 \alpha}~\left[~\frac{2m}{\hbar^2} 
~\mu~\right]^{3/2}\;.
\end{equation}
But from Eq.~(\ref{pond}), $\alpha$ is proportional to 
$\sqrt{\mu}$, yielding the correct linear depence between the density 
$\rho$ and $\mu$ as in BEC. We then obtain the same sound velocity $u_s$ as 
in BEC : 
\beq
u_s^2 = \frac{\rho_a}{m} \frac{\partial \mu}{\partial \rho_a} = 
\frac{\mu}{m}.
\eeq
By contrast, $u_s^2=2\mu/3m$ in a free Fermi gas with constant 
$\alpha$.
By setting $k_F^2 =\frac{2m\mu}{\hbar^2}$, we see from Eq.~(\ref{pond}) 
that $\alpha$ is proportional to the dimensionless quantity $a k_F$. 
This results in the occupancy factor $1/\alpha$ being 
\begin{equation}
{1\over \alpha}=\sqrt{{7\over {12}}} {\lambda_F\over {2a}}\,.
\label{gowda}
\end{equation}
Since the de Broglie wave length $\lambda_F=2\pi/k_F$ may be regarded as the 
resolving power, we may interpret the degeneracy described by the
parameter $1/\alpha $ as the collective number of the atoms which can be
accommodated within a wave length. The quantum mechanical wave
functions of all these atoms overlap substantially over a wave length. 
Thus a single ``fermion''  in our picture is as if made of $1/\alpha $ 
number of mock-fermions. It is also easy to check that 
\begin{equation}
k_F^{-1}=(8 \pi a \rho)^{-1/2}=\xi\;,
\end{equation} 
where $\xi$ is the coherence length of the bosons in the 
condensate~\cite{stringari}.
Using the parameters of the experiment~\cite{anderson}, 
it is straight-forward to get a numerical estimate of this 
collectivity. For example, for $N=5000$, we find $1/\alpha = 140$. Note 
that in our picture the mock-fermions are essentially free apart from the 
confinement in the plane and over a length in the z-direction. The single 
fermion excitations in this picture may now be described as the 
excitation of $1/\alpha$  mock-fermions. This has interesting 
consequences in computing the excitation energies of the system.

As an immediate application we may consider the vortex states 
discussed by Dalfalvo and Stringari\cite{stringari}. 
In our Fermi gas picture, we may make an estimate for a vortex excitation 
in the static ``shell model'' picture. For making this estimate, 
consider the ground state in which mock-fermions multiply occupy the states 
of the two-dimensional harmonic oscillator upto the Fermi energy. 
Since the density profile of a vortex has 
a hole in the centre, we may simulate it by promoting all the s-state 
mock-fermions out of the Fermi sea to states with $l\neq 0$.
Each of these 
s-states contain $1/\alpha$ mock-fermions in our picture. 
Keeping $N$ fixed, the number of $l=0$ states below the 
chemical potential $\mu$ is given by $\mu/(2\hbar\omega)$ for large 
$\mu$. Since the degeneracy of the harmonic oscillator state just above the 
the Fermi energy is $(\mu +1)/\hbar \omega_{\perp}$, it is possible to promote 
all the s-state particles to this state (upto a maximum angular momentum).
The excitation energy per particle in the presence of a vortex is then 
given by, 
\beq
\Delta E/N=
\frac{1}{N\alpha}(\frac{\mu}{2\hbar\omega_{\perp}})^2 \hbar \omega_{\perp}. 
\label{punch}
\eeq
This should be compared with the energy  $\hbar \Omega_c$ calculated by 
Dalfovo and Stringari~\cite{stringari} for $\kappa=1$, where $\Omega_c$ 
is the critical 
angular velocity and $\hbar \kappa$ the angular momentum of the vortex.
In Fig.1, we display the result of our estimate for $\Omega_c$, obtained from 
Eq.~(\ref{punch}), with the graph given in~\cite{stringari} for the same. 
The agreement for large $N$ is good between our calculations 
and those obtained by solving the non-linear Schr\"{o}dinger equation. 
For comparison we have also displayed the calculated values of $\Omega_c$ 
obtained in the large $N$ limit by Baym and Pethick\cite{baym}. While the 
power law 
dependence on the chemical potential is the same in our calculation 
and in \cite{baym}, we cannot get 
the dependence on the logarithm of the chemical potential in the static 
shell model type of analysis. In our naive model, the 
vortex energy is insensitive to the vortex angular momentum $\hbar \kappa $ 
upto a maximum.

~~~~~As pointed out in the introduction, the mock-fermion picture also 
predicts  collective particle-hole  excitations from the filled Fermi level to 
an excited state, where all ${1\over \alpha}$ mock fermions of a 
given angular momentum state are excited together. Since the adjacent shells 
of the harmonic oscillator have states of opposite parity, excitations to the 
next higher state of the ${1\over \alpha}$ mock fermions whould entail a huge 
change in angular momentum. Such 
high-spin collective excitaions are unlikely. There could be, however, 
collective excitations of energy
${2\over \alpha} \hbar \omega_{\perp}$ and zero angular momentum. 
Both the above types of excitations are of a novel collective type, and 
involve a hundred or more mock-fermions. As such, they are much larger 
energy 
excited states than the ones predicted and analysed  by Stringari~\cite{str} 
recently.

~~~~To conclude, the Bose-Einstein condensate of interacting bosons is 
described in this paper by a noninteracting Fermi gas. 
We call them mock fermions since they obey a generalised Pauli principle 
with the occupancy of a state $i$ given by 
\begin{eqnarray}
n_i&=&{1\over \alpha},~\epsilon_i < \mu \; ; \nonumber \\
   &=&0,~\epsilon_i > \mu \;.
\label{cash}
\end{eqnarray}
This occupancy factor is directly proportional to the coherance length 
$\xi=(8\pi a \rho)^{-1/2}$ divided by the ``size'' $2a$ of the atom. 
The collectivity is generated by the combined motion of $1/\alpha$ mock 
fermions. 
In passing, it is worth mentioning that an occupancy factor of 
the type~(\ref{cash}) at $T=0$ arises naturally in systems that obey the 
so-called Haldane 
statistics~\cite{haldane}. It has also been noted that in such 
systems that the single particle density is scaled by the occupancy 
factor~\cite{dsen}. 
Irrespective of this aspect, the 
present analysis of the bosonic  
condensate in terms of mock-fermions is intersting in its own 
right, although one should only expect it to hold for global properties, 
and not for the analysis of correlations.

We gratefully acknowledge comments from Diptiman Sen and discussions with
Jim Waddington.  One of us (MVNM) thanks the Department of Physics and
Astronomy, McMaster University for hospitality. This research was
supported by National Science and Engineering Council of Canada.

\begin{figure}
\caption{
The critical frequency $\Omega_c$ (in units of $\omega_{\perp}$) is 
shown as a function of the particle 
number in the fermion-like picture (solid line). The dotted line 
is based on the graph given by Dalfovo and Stringari[6]. The 
dashed line is the large $N$ prediction[5]. }
\label{Fig.1}
\end{figure}

\end{document}